\documentclass[prb,showpacs,twocolumn,aps]{revtex4}
\usepackage{graphicx}
\newcommand{\be}{\begin{equation}}
\newcommand{\ee}{\end{equation}}
\newcommand{\bea}{\begin{eqnarray}}
\newcommand{\eea}{\end{eqnarray}}

\newcommand{\etal}{{\it et al.} }
\begin{document}
\title{Transport properties in the $d$-density wave state: 
Wiedemann-Franz law} 
\author{Wonkee Kim and J. P. Carbotte}
\affiliation{Department of Physics and Astronomy,
McMaster University, Hamilton,
Ontario, Canada, L8S~4M1}
\begin{abstract}
We study the Wiedemann-Franz (WF) law in the $d$-density wave (DDW) model.
Even though the opening of the DDW gap $(W_{0})$  profoundly modifies the
electronic density of states and makes it dependent on energy, 
the value of the WF ratio at zero
temperature $(T=0)$ remains unchanged. 
However, neither electrical nor thermal conductivity
display universal behavior. For finite temperature, with $T$ greater than
the value of the impurity scattering rate at zero frequency $\gamma(0)$
{\it i.e.} $\gamma(0)<T\ll W_{0}$, the usual WF ratio is
obtained only in the weak scattering limit. For strong scattering
there are large violations of the WF law.
\end{abstract}
\maketitle

In a recent paper Hill \etal\cite{hill} have observed large violation of the
Wiedemann-Franz (WF) law in (Pr,Ce)$_{2}$CuO$_{2}$ driven into the normal state 
through the application of a $13$ Tesla magnetic field. At very low 
temperature $T<0.2K$, the thermal conductivity is found to be much less than
the value estimated from the D.C. conductivity.
Above $0.3K$ the opposite holds. This observation suggests that an exotic
state of matter may exist in the normal state of (Pr,Ce)$_{2}$CuO$_{2}$.
Hill \etal consider spin-charge separation as one possibility.

Recently $d$-density wave (DDW) order has received considerable attention
\cite{chakravarty1,chakravarty2,dhlee,zhu,kim} as a possible exotic
state of matter with a pseudogap which breaks time reversal symmetry because
it introduces bond current with attendant small orbital magnetic moments.
The pseudogap has $d$-wave symmetry. This is the symmetry 
observed in studies of the variation of 
the leading edge of the electron spectral 
density by angle-resolved photoemission spectroscopy,
\cite{arpes} as a function of angle in the Brillouin zone in the normal state
of underdoped cuprates. A pseudogap with $d$-wave symmetry implies
important energy dependence of the quasiparticles density of states
(DOS) at the Fermi surface (FS). Energy dependence in
the DOS leads to impurity scattering rates that also depend on energy
and
the applicability of the usual WF law is no longer
guaranteed.

In this paper we consider the WF law within the DDW model.
As yet, this model
has not been shown to apply to the pseudogap regime
of the cuprates. Here we take the point of view
that nevertheless
it can serve to understand, in this concrete case, how energy dependence 
in the DOS can alter the WF law.

In the DDW state, the gap with magnitude $W_{0}$ has $d$-wave
symmetry and opens up at the antiferromagnetic Brillouin zone of the CuO$_{2}$
plane. Away from half filling, in the underdoped regime, the FS falls
at the chemical potential $\mu$ (which would be zero at half filling) and
we assume that $|\mu|\ll W_{0}$. Provided that the effective impurity
scattering rate and temperature are also small as compared with $W_{0}$,
a nodal approximation\cite{durst} can be used to 
describe the electric 
as well as
the thermal conductivity. 

We consider a tight binding energy dispersion as a function of
momentum ${\bf k}$ of the form:
$\epsilon_{\bf k}=-2t_{0}[\cos(k_{x})+\cos(k_{y})]$, 
where $t_{0}$ is the in-plane
hopping amplitude.
At half filling the FS  coincides with
the antiferromagnetic boundary where the DDW gap
$W_{\bf k}=(W_{0}/2)[\cos(k_{x})-\cos(k_{y})]$ opens up
with amplitude $W_{0}$. Most properties of the DDW state are determined by
the nesting vector
${\bf Q}=(\pi,\pi)$, for example $\epsilon_{{\bf k}+{\bf Q}}=
-\epsilon_{\bf k}$ and $W_{{\bf k}+{\bf Q}}=-W_{\bf k}$. 
See Ref.\cite{zhu,kim} for detailed properties.

We begin with the Hamiltonian
\bea
H = && \int d{\bf x}\psi^{\dagger}_{\alpha}({\bf x})
\left(-\frac{\nabla^{2}}{2m}\right) \psi_{\alpha}({\bf x})
\nonumber\\
+&& \frac{1}{2} \int d{\bf x}d{\bf y}\psi^{\dagger}_{\alpha}({\bf x})
\psi^{\dagger}_{\beta}({\bf y}) V({\bf x} - {\bf y})
\psi_{\beta}({\bf y}) \psi_{\alpha}({\bf x}),
\label{hamiltoinan}
\eea
where $\psi^{\dagger}_{\alpha}({\bf x})$ creates an electron of spin $\alpha$
at ${\bf x}$ and $V({\bf x} - {\bf y})$ is the electron-electron
interaction. A spin summation is implied. Using the 
definition
of the DDW gap in momentum space:
\be
i W_{\bf k}=-\sum_{k'}V_{{\bf k}-{\bf k}'}\langle
C^{\dagger}_{{\bf k}'+{\bf Q}\alpha}(\omega')C_{{\bf k}'\alpha}(\omega')
\rangle
\label{ddwgap}
\ee
one can obtain the mean field Hamiltonian of the DDW state.

Let us consider the real part of the electrical conductivity 
$\sigma(\Omega)$. 
In the long wavelength
limit $({\bf q}\rightarrow0)$ the current operator in momentum space
${\bf j}^{e}(0,\Omega)$ is
\be
{\bf j}^{e}(0,\Omega) = -e \sum_{k} {\bf v}_{\bf k}
{\hat C}^{\dagger}_{\bf k}(\omega)
{\hat\tau}_{3}{\hat C}_{\bf k}(\omega+\Omega)\;,
\ee
where ${\bf v}_{\bf k}=\partial\epsilon_{k}/\partial {\bf k}$ and
${\hat C}^{\dagger}_{\bf k}(\omega)=
\left(C^{\dagger}_{{\bf k}\uparrow}(\omega),\;
C^{\dagger}_{{\bf k}+{\bf Q}\uparrow}(\omega)\right)$.
Note that we use four vector notation: $k=({\bf k},\omega)$ and
$\sum_{k}=\sum_{\bf k}\sum_{\omega}$. 
From the current-current correlation $\Pi(0,i\Omega)$, we have
$\sigma(\Omega)=-\frac{1}{\Omega}\mbox{Im}\Pi_{ret}(\Omega)$, where
\be
\Pi(0,i\Omega)=e^{2}\sum_{k}v^{2}_{\bf k}\mbox{Tr}
\left[{\hat G}({\bf k},i{\tilde\omega}+i\Omega){\hat\tau}_{3}
{\hat G}({\bf k},i{\tilde\omega}){\hat\tau}_{3}\right]
\label{Pi_DDW}
\ee
Here the matrix Green's function ${\hat G}({\bf k},i{\tilde\omega})$ is
\be
{\hat G}({\bf k},i{\tilde\omega})=
\frac{(i{\tilde\omega}+\mu){\hat\tau}_{0}+W_{\bf k}{\hat\tau}_{2}+
\epsilon_{\bf k}{\hat\tau}_{3}}
{(i{\tilde\omega}+\mu)^{2}-E^{2}_{\bf k}}
\ee
with $i{\tilde\omega}=i\omega-\Sigma_{0}(i{\tilde\omega})$
and $E_{\bf k}=\sqrt{\epsilon^{2}_{\bf k}+W^{2}_{\bf k}}$.
It has been assumed that the other components of 
the self energy can be absorbed into
$\epsilon_{\bf k}$ and $W_{\bf k}$.\cite{durst}
It is useful to introduce the spectral functions 
$A_{ij}({\bf k},\omega)=-2\mbox{Im}G_{ij}({\bf k},\omega+i\delta)$,
for example, $A_{11}({\bf k},\omega)=
\frac{2\gamma(\omega)|u_{\bf k}|^{2}}
{\left[(\omega+\mu-E_{\bf k})^{2}+\gamma(\omega)^{2}\right]}
+\left(u_{\bf k}\rightarrow v_{\bf k}, E_{\bf k}\rightarrow -E_{\bf k}\right)$,
and
$A_{12}({\bf k},\omega)=-u_{\bf k}v_{\bf k}\left\{
\frac{2\gamma(\omega)}
{\left[(\omega+\mu+E_{\bf k})^{2}+\gamma(\omega)^{2}\right]}
-\left(E_{\bf k}\rightarrow -E_{\bf k}\right)
\right\}$, where $\gamma(\omega)=-\mbox{Im}\Sigma_{0,ret}(i{\tilde\omega})$, 
$u_{\bf k}=\sqrt{\frac{1}{2}(1+\epsilon_{\bf k}/E_{\bf k})}$, and
$v_{\bf k}=i\sqrt{\frac{1}{2}(1-\epsilon_{\bf k}/E_{\bf k}})$.
Now we obtain the D.C. conductivity $\sigma(T,\Omega=0)$ as
\begin{small}
\be
\sigma(T,0)=e^{2}\sum_{\bf k}v^{2}_{\bf k}\int\frac{d\omega}{2\pi}
\left(-\frac{\partial f}{\partial\omega}\right)
\left[A_{11}({\bf k},\omega)^{2}-
|A_{12}({\bf k},\omega)|^{2}\right]
\ee
\end{small}
where $f(\omega)$ is the Fermi function.
In the nodal approximation $\sum_{\bf k}\rightarrow
\frac{4}{v_{f}v_{g}}\int\frac{pdpd\theta}{(2\pi)^{2}}$,
$\epsilon_{\bf k}=p\cos(\theta)$ and
$W_{\bf k}=p\sin(\theta)$. Then at $T=0$ we obtain
$
\sigma(0,0)=\frac{e^{2}}{\pi^{2}}
\left(\frac{v_{f}}{v_{g}}\right){\cal A}(0)\;,
$
where
\be
{\cal A}(\omega)=
\left[1+\left(\frac{\omega+\mu}{\gamma(\omega)}+
\frac{\gamma(\omega)}{\omega+\mu}\right)
\arctan\frac{\omega+\mu}{\gamma(\omega)}
\right]\;.
\label{A_ddw}
\ee
$\sigma(0,0)$ depends only on $\gamma(0)$ because only the
zero frequency limit of ${\cal A}(\omega)$ enters at $T=0$.

This result shows that $\sigma(0,0)$ depends not only on the chemical potential
$\mu$ (and so on the filling) but also on the scattering rate
$\gamma(0)$. This is to be contrasted with the well-known universal
value of the DC conductivity for the DSC: $\sigma_{sc}(0,0)=2\frac{e^{2}}
{\pi^{2}}(v_{f}/v_{sc,g})$, where $v_{sc,g}$ is the DSC gap velocity.
For the DDW case a universal value is obtained only in the case when
$\mu\rightarrow0$, which corresponds to half filling. In this limit
$\sigma(0,0)$  reduces precisely to $\sigma_{sc}(0,0)$ for
the DSC with $W_{0}$ playing the role of DSC gap $(\Delta_{0})$. 
We see that it is because the DDW gap develops
at the antiferromagnetic boundary rather than at the FS which is shifted 
by the chemical
potential, which leads to the absence of universal behavior.

It is instructive to contrast the DDW case with the DSC case in a more 
formal way. The charge current has the form for the DSC
\be
{\bf j}^{e}_{sc}(0,\Omega)=-e \sum_{k} {\bf v}_{\bf k}
{\hat \psi}^{\dagger}_{\bf k}(\omega)
{\hat\psi}_{\bf k}(\omega+\Omega)
\ee
where
${\hat \psi}^{\dagger}_{\bf k}(\omega)=
\left(C^{\dagger}_{\bf k\uparrow}(\omega),\;
C_{-{\bf k}\downarrow}(\omega)\right)$. This leads to the current-current
correlation
\be
\Pi_{sc}(0,i\Omega)=e^{2}\sum_{k}v^{2}_{\bf k}\mbox{Tr}
\left[{\hat G}_{sc}({\bf k},i{\tilde\omega}+i\Omega)
{\hat G}_{sc}({\bf k},i{\tilde\omega})\right]
\ee
which is to be contrast with Eq.~(\ref{Pi_DDW}). The matrix Green's function
is:
\be
{\hat G}_{sc}({\bf k},i{\tilde\omega})=
\frac{i{\tilde\omega}{\hat\tau}_{0}+\Delta_{\bf k}{\hat\tau}_{1}+
(\epsilon_{\bf k}-\mu){\hat\tau}_{3}}
{(i{\tilde\omega})^{2}-(\epsilon_{\bf k}-\mu)^{2}-\Delta^{2}_{\bf k}}
\ee
Note the differences between ${\hat G}_{sc}$ for the DSC and
${\hat G}$ for the DDW.
Using the spectral function $A({\bf k},\omega)=-\mbox{Im}
G({\bf k},\omega+i\delta)$ and $B({\bf k},\omega)=-\mbox{Im}
F({\bf k},\omega+i\delta)$, 
where $F({\bf k},\omega)$ is the anomalous Green's function,
the D.C. conductivity becomes
\be
\sigma_{sc}(T,0)=e^{2}\sum_{\bf k}v^{2}_{\bf k}\int\frac{d\omega}{2\pi}
\left(-\frac{\partial f}{\partial\omega}\right)
\left[A({\bf k},\omega)^{2}+
B({\bf k},\omega)^{2}\right]
\ee
At $T=0$, we obtain 
$
\sigma_{sc}(0,0)=\frac{e^{2}}{\pi^{2}}
\left(\frac{v_{f}}{v_{sc,g}}\right){\cal A}_{sc}(0)\;,
$
where
\be
{\cal A}_{sc}(\omega)=
2\left[1+\frac{\omega}{\gamma(\omega)}
\arctan\frac{\omega}{\gamma(\omega)}
\right]\;.
\label{A_dsc}
\ee

Next we consider the case of finite $T$ in the range
$\gamma(0)<T\ll W_{0}$. In this case
$i{\tilde\omega}\simeq i\omega-\Sigma_{0}(i\omega)$, namely, $\omega$
can be used in the evaluation of $\Sigma_{0}$ to a good approximation.
\cite{hirschfeld} Then
$\Sigma_{0,ret}(i\omega)=\frac{\Gamma G_{0}}{c^{2}-G^{2}_{0}}$, where
$G_{0}=
\frac{1}{\pi N_{0}}
\sum_{\bf k}\frac{\omega+\mu}{(\omega+\mu)^{2}-E^{2}_{\bf k}}$
with $N_{0}$ being the DOS at the FS,
$\Gamma$ is a scattering rate
proportional to the impurity concentration, and $c$ is
the inverse of the impurity potential. For the Born limit $c\gg 1$ while
in the unitary limit $c\rightarrow0$. Applying the nodal approximation,
one obtains $G_{0}=\frac{2}{\pi^{2}N_{0}v_{f}v_{g}}\left[
-i\frac{\pi}{2}(\omega+\mu)+(\omega+\mu)\ln\left(\frac{|\omega+\mu|}
{W_{0}}\right)\right]$. Thus for the Born limit
we get $\gamma(\omega)=\gamma_{0}
\frac{\omega+\mu}{W_{0}}$, where $\gamma_{0}\approx\Gamma/c^{2}$. 
For the unitary limit we have instead
$\gamma(\omega)=\gamma_{u}\left(\frac{W_{0}}{\omega+\mu}\right)
\ln^{-2}\left(\frac{W_{0}}{|\omega+\mu|}\right)$ where
$\gamma_{U}\approx\pi^{2}\Gamma/4$.
These results for $\gamma(\omega)$ 
parallel the well-known results for the DSC, which are
recovered when $\mu=0$ with the DDW gap replaced with
the DSC gap. The most important feature of impurity scattering for
our consideration of transport properties is that $\gamma(\omega)$
acquires a frequency dependence and this leads to a violation with $T$
of the WF law as we will see soon. For the Born limit a remarkable
simplification for ${\cal A}(\omega)$ occurs; namely, ${\cal A}(\omega)$
becomes independent of frequency and this leads directly to no violation of
the WF law. With ${\cal A}(\omega)=1+\left(\frac{W_{0}}{\gamma_{0}}+
\frac{\gamma_{0}}{W_{0}}\right)\arctan\left(\frac{W_{0}}{\gamma_{0}}\right)
\simeq\frac{\pi}{2}\frac{W_{0}}{\gamma_{0}}$ because $\gamma_{0}\ll W_{0}$
and it follows immediately that
\be
\sigma(0,T\ll W_{0})\simeq
\frac{e^{2}}{2\pi}\left(\frac{v_{f}}{v_{g}}\right)
\frac{W_{0}}{\gamma_{0}}
\label{sigma_ddw_T}
\ee
which is temperature-independent. For the DSC in the same limit
$\sigma_{sc}(0,T\ll\Delta_{0})\simeq (e^{2}/\pi)
(v_{f}/v_{sc,g})(\Delta_{0}/\gamma_{0})$. It is larger than
the DDW results by a factor of two if $W_{0}=\Delta_{0}$. 
The difference is traced
to the fact that 
${\cal A}_{sc}(\omega)\simeq 2\left[(\pi/2)\Delta_{0}/\gamma_{0}
\right]$. This serves to illustrate that DSC and DDW order do not generally
give the same answers. 
This is expected since
in one case there is Cooper pair condensation while in the other
there is none.

We next consider heat transport in the DDW state since the WF law is a 
statement about the ratio of the thermal to electrical conductivity.
The heat current ${\bf j}^{h}({\bf x})$ can be calculated from
the continuity equation: 
$\dot{\cal H}({\bf x}) + \nabla \cdot {\bf j}^{h}({\bf x}) = 0$,
where is $\cal H$ is the Hamiltonian density of Eq.~(\ref{hamiltoinan}). 
Define
${\bf j}^{h}={\bf j}^{h}_{f}+{\bf j}^{h}_{g}$, one can show that
\be
{\bf j}^{h}_{f}({\bf x})= -\frac{1}{2m} \left[ \dot{\psi}_{\alpha}^{\dagger}(x)
\nabla\psi_{\alpha}(x) + \nabla\psi_{\alpha}^{\dagger}(x)
\dot{\psi}_{\alpha}(x) \right]
\ee
and
\bea
\nabla\cdot{\bf j}^{h}_{g}({\bf x}) =&& \frac{1}{2}\int d{\bf y}
V({\bf y} - {\bf x}) 
\Bigl[\dot{\psi}_{\alpha}^{\dagger}(x) \psi_{\beta}^{\dagger}(y)
\psi_{\beta}(y) \psi_{\alpha}(x)
\nonumber\\
-&&\psi_{\alpha}^{\dagger}(x) \dot{\psi}_{\beta}^{\dagger}(y)
\psi_{\beta}(y) \psi_{\alpha}(x)+h.c.\Bigr]
\eea
In momentum space, as ${\bf q}\rightarrow0$,
\be
{\bf j}^{h}_{f}(0,\Omega) = \sum_{k}\left(\omega+\frac{\Omega}{2}\right)
{\bf v}_{\bf k}
{\hat C}^{\dagger}_{\bf k}(\omega)
{\hat\tau}_{3}{\hat C}_{\bf k}(\omega+\Omega)
\ee
Applying mean field theory
and keeping terms relevant only to the DDW order, we find
$
i{\bf q}\cdot {\bf j}^{h}_{g}({\bf q},\Omega)=\frac{1}{2}
(X_{q}+X^{*}_{-q}-Y_{q}-Y^{*}_{-q})
$
where
$
X_{q}=\sum_{k}\omega W_{k+q}C^{\dagger}_{{\bf k}+{\bf Q}\alpha}(\omega)
C_{{\bf k}+{\bf q}\alpha}(\omega+\Omega)
$ and
$
Y_{q}=X_{g}\left(W_{k+q}\rightarrow W_{q}\right)
$
with a definition of the DDW gap Eq.~(\ref{ddwgap}).
Now we obtain
\be
{\bf j}^{h}_{g}(0,\Omega)=-i\sum_{k}\left(\omega+\frac{\Omega}{2}\right)
{\bf v}_{g}
{\hat C}^{\dagger}_{{\bf k}+{\bf Q}}(\omega)
{\hat\tau}_{3}{\hat C}_{\bf k}(\omega+\Omega)
\ee
where ${\bf v}_{g}=\partial W_{k}/\partial {\bf k}$.
Therefore,
the heat current becomes
\bea
{\bf j}^{h}(0,\Omega)=&&\sum_{k}\left(\omega+\frac{\Omega}{2}\right)
\Bigl[
{\bf v}_{\bf k}
{\hat C}^{\dagger}_{\bf k}(\omega)
{\hat\tau}_{3}{\hat C}_{\bf k}(\omega+\Omega)
\nonumber\\
&&-i {\bf v}_{g}{\hat C}^{\dagger}_{{\bf k}+{\bf Q}}(\omega)
{\hat\tau}_{3}{\hat C}_{\bf k}(\omega+\Omega)\Bigr]
\eea
Note that
we assume $\partial {\bf v}_{g}/\partial t=0$ so that we neglect
the extra terms which depend on the time derivative of
the gap velocity.

The thermal conductivity $\kappa({\bf q}=0,\Omega)$ 
follows from the Kubo
formula for the heat current-current correlation:
$\frac{\kappa(\Omega)}{T}=-\frac{1}{T^{2}\Omega}
\mbox{Im}\Pi^{\kappa}_{ret}(\Omega)$,
where
\begin{small}
\bea
&&\Pi^{\kappa}(0,i\Omega)=\sum_{k}\left(\omega+\frac{\Omega}{2}\right)^{2}
v^{2}_{\bf k}\mbox{Tr}
\left[{\hat G}({\bf k},i{\tilde\omega}+i\Omega){\hat\tau}_{3}
{\hat G}({\bf k},i{\tilde\omega}){\hat\tau}_{3}\right]
\nonumber\\
&&+\sum_{k}\left(\omega+\frac{\Omega}{2}\right)^{2}
v^{2}_{g}\mbox{Tr}
\left[{\hat G}({\bf k},i{\tilde\omega}+i\Omega){\hat\tau}_{3}
{\hat G}({\bf k}+{\bf Q},i{\tilde\omega}){\hat\tau}_{3}\right]
\eea
\end{small}
Again making use of the spectral functions, we obtain
\bea
\frac{\kappa(0)}{T}=&&\frac{1}{T^{2}}\sum_{k}
\int \frac{d\omega}{2\pi}\omega^{2}
\left(-\frac{\partial f}{\partial\omega}\right)
\left(v^{2}_{f}+v^{2}_{g}\right)
\nonumber\\
\times
&&\left[A_{11}({\bf k},\omega)^{2}-
|A_{12}({\bf k},\omega)|^{2}\right]
\label{thermal_ddw}
\eea
Before  proceeding further it is of interest to contrast our DDW derivations
with the DSC case. For the DSC the heat current is
\bea
{\bf j}^{h}_{sc}(0,\Omega)=&&\sum_{k}\left(\omega+\frac{\Omega}{2}\right)
\Bigl[
{\bf v}_{\bf k}
{\hat \psi}^{\dagger}_{\bf k}(\omega)
{\hat\tau}_{3}{\hat \psi}_{\bf k}(\omega+\Omega)
\nonumber\\
&&-{\bf v}_{g,sc}{\hat \psi}^{\dagger}_{\bf k}(\omega)
{\hat\tau}_{1}{\hat \psi}_{\bf k}(\omega+\Omega)\Bigr]
\eea
Thus the heat current-current correlation is
\begin{small}
\bea
&&\Pi^{\kappa}_{sc}(0,i\Omega)=
\sum_{k}\left(\omega+\frac{\Omega}{2}\right)^{2}
v^{2}_{\bf k}\mbox{Tr}
\left[{\hat G}_{sc}({\bf k},i{\tilde\omega}+i\Omega){\hat\tau}_{3}
{\hat G}_{sc}({\bf k},i{\tilde\omega}){\hat\tau}_{3}\right]
\nonumber\\
&&+\sum_{k}\left(\omega+\frac{\Omega}{2}\right)^{2}
v^{2}_{sc,g}\mbox{Tr}
\left[{\hat G}_{sc}({\bf k},i{\tilde\omega}+i\Omega){\hat\tau}_{1}
{\hat G}_{sc}({\bf k},i{\tilde\omega}){\hat\tau}_{1}\right]
\eea
\end{small}
When the spectral functions are introduced, we arrive at
\bea
\frac{\kappa_{sc}(T)}{T}=&&\frac{1}{T^{2}}
\sum_{\bf k}\int\frac{d\omega}{2\pi}
\omega^{2}\left(-\frac{\partial f}{\partial\omega}\right)
\nonumber\\
&&\times
\left(v^{2}_{f}+v^{2}_{sc,g}\right)
\left[ A({\bf k},\omega)^{2}-B({\bf k},\omega)^{2}\right]
\eea

Applying the nodal approximation to Eq.(\ref{thermal_ddw}), we obtain
\be
\frac{\kappa(T)}{T}=\frac{1}{T^{2}}
\sum_{\bf k}\int d\omega\frac{\omega^{2}}{\pi^{2}}
\left(-\frac{\partial f}{\partial\omega}\right)
\left[\frac{v_{f}}{v_{g}}+\frac{v_{g}}{v_{f}}\right]{\cal A}(\omega)
\ee
and exactly the same result holds for the DSC 
with $\mu\rightarrow0$ and $v_{g}\rightarrow v_{sc,g}$.
As $T\rightarrow0$, $\kappa/T=\frac{1}{3}
\left(\frac{v_{f}}{v_{g}}+\frac{v_{g}}{v_{f}}\right){\cal A}(0)$ for the DDW
while $\kappa_{sc}/T=\frac{2}{3}
\left(\frac{v_{f}}{v_{sc,g}}+\frac{v_{sc,g}}{v_{f}}\right)$ for the DSC.
In this case
$\kappa_{sc}/T$ is universal and does not depend on impurity
scattering. In contrast $\kappa/T$ for the DDW has a dependence
on $\gamma(0)$ as well as on doping through the chemical potential.
However, for the Lorenz number $L=\kappa(T)/[T\sigma(0,0)]$ the scattering rate 
drops out and we find
$
L_{0}=\frac{\pi^{2}}{3e^{2}}\left[1+\left(\frac{v_{g}}{v_{f}}\right)^{2}\right]
$.
This shows that the WF law is obeyed at $T=0$ in the DDW state
and the DSC case $(v_{g}\rightarrow v_{sc,g})$ and its value
differs from the conventional one only by a very small correction 
of order $(v_{g}/v_{f})^{2}$ due to
a $d$-wave symmetry of the gap.

A very similar result can be obtained in the case $\gamma(0)<T\ll W_{0}$.
In this regime we have already seen
${\cal A}(\omega)\simeq\frac{\pi}{2}\frac{W_{0}}{\gamma_{0}}$ 
for the Born limit so
$
\frac{\kappa(T)}{T}=\frac{\pi}{6}\left(\frac{v_{f}}{v_{g}}+
\frac{v_{g}}{v_{f}}\right)\frac{W_{0}}{\gamma_{0}}
$.
For the DSC, $W_{0}\rightarrow\Delta_{0}$ and $v_{g}\rightarrow v_{sc,g}$.
Because of Eq.(\ref{sigma_ddw_T}) for the DDW the Lorenz number reduces
to the conventional value: $L=\pi^{2}/(3e^{2})$  for $\gamma(0)<T\ll W_{0}$.
But for the DSC we arrive instead at the remarkable result that
$L_{sc}=\pi^{2}/(6e^{2})$, a reduction of a factor of two. While we 
obtain this results analytically, Graf \etal \cite{graf}
have calculated $L_{sc}$ numerically and their work serves as a
numerical verification of our result.
Since the temperature scale for
which this happens is $\gamma(0)\ll T$, in the clean limit 
this switch-over from $L_{0}$ to $L_{0}/2$ can happen
at extremely low $T$. 
In sharp contrast with the DSC, in the DDW case there is 
no change in the Lorenz number in the Born limit.

\begin{figure}[tp]
\begin{center}
\includegraphics[width=\linewidth,clip]{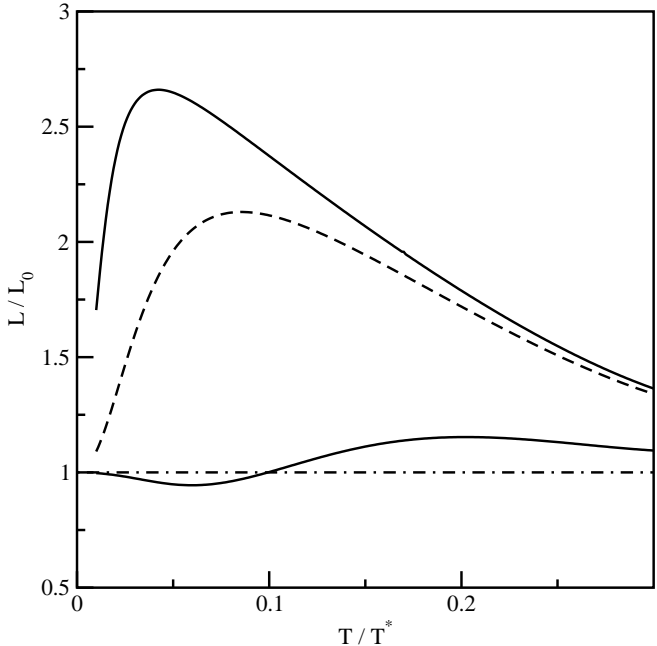}
\caption{
The normalized Lorenz number as a function of temperature $(T)$.
The dash-dotted curve is the result for the Born limit.
Other  curves are for the unitary limit.
For the upper solid curve $\gamma_{U}/W_{0}=0.001$ and $\mu/W_{0}=-0.01$.
The dashed curve is for the same chemical potential but for
$\gamma_{U}/W_{0}=0.01$. The lower solid curve
is for $\gamma_{U}/W_{0}=0.001$ but now $\mu/W_{0}=-0.15$.
}
\end{center}
\end{figure}
It is not possible to obtain analytic results for the unitary limit. In general
the Lorenz number $L(T)/L(0)$ is written as
\be
\frac{L(T)}{L_{0}}=
\frac{3}{\pi^{2}}
\frac{\int d\omega\left(\frac{\omega}{T}\right)^{2}
\left(-\frac{\partial f}{\partial\omega}\right){\cal A}(\omega)}
{\int d\omega
\left(-\frac{\partial f}{\partial\omega}\right){\cal A}(\omega)}\;.
\ee
As we mentioned earlier, in the unitary limit 
$\gamma(\omega)=\gamma_{U}\left(\frac{W_{0}}{\omega+\mu}\right)
\ln^{-2}\left(\frac{W_{0}}{|\omega+\mu|}\right)$. Numerical results
are presented in Fig.~1. We show results for four different cases which serve
to illustrate what is possible. The upper solid curve is for
$\gamma_{U}/W_{0}=0.001$ and $\mu/W_{0}=-0.01$ which shows a large peak around
$T/T^{*}=0.05$. We have taken $T^{*}$ to be given by its mean field
value: $W_{0}/T^{*}=2.14$ as in the DSC case. 
A very large positive violation of the WF law is seen. We need to point out,
however, that while we have not shown $\sigma(T)$ and $\kappa(T)/T$
individually, in this case they both show large variations with $T$ 
reflecting the important frequency variation of $\gamma(\omega)$
for the unitary limit, which is not compensated for by the explicit variation
of ${\cal A}(\omega)$ in Eq.(\ref{A_ddw}). For the Born limit
an exact compensation takes place so that ${\cal A}(\omega)$ turns out to be
a constant. This leads to the usual WF law with no $T$ dependence, which is
shown as a dash-dotted line in Fig.~1. 
For the dashed curve $\gamma_{U}/W_{0}=0.01$
and $\mu/W_{0}=-0.01$. Increasing $\gamma_{U}$ makes the deviations from
the conventional Lorenz number
smaller.  
The same effect is obtained when $|\mu|$ is increased, effectively
pushing the FS further away from the zero in DOS. 
The second solid curve has $\gamma_{U}/W_{0}=0.001$ but now
$\mu/W_{0}=-0.15$, away from half filling.
Now the deviation from the conventional Lorenz number can be negative
as well as positive depending on $T$ but the amplitude of the
violation is small because the DDW gap becomes less effective at changing 
the
DOS near the FS. (Note that $|\mu|\ll W_{0}$
for the validity of the nodal approximation.)

Our main conclusions are as follows. At $T=0$, only the zero frequency
limit of the imaginary part of the self-energy enters into 
the calculation of the electrical and
thermal conductivity and the conventional Wiedemann-Franz (WF) 
law is recovered.
In contrast with what is found for a $d$-wave superconductor (DSC),
for a $d$-density wave (DDW) state,
neither electrical nor thermal conductivity show universal behavior.
Each depends on the impurity scattering rate.
But this dependence is the same and
cancels from the Lorenz
number as $T\rightarrow0$. We were also able to
obtain analytic results for low but finite temperature.
In this case we found no change in the WF law for the Born limit 
even though
the Lorenz number is reduced by a factor of two
from its conventional value for the DSC. 
For the unitary limit, however,
the Lorenz number increases rapidly at low temperature 
on a scale set by the zero scattering rate $\gamma(0)$. In a case
considered it rises above $2.5$ around $T/T^{*}\simeq0.05$ 
and then acquires a more moderate temperature 
variation. This case corresponds to the chemical potential $(\mu)$ 
small compared to the DDW gap.
When $|\mu|$ is increased sufficiently, the Lorenz number 
becomes approximately equal to 
its conventional value and its temperature dependence 
is small.
It is important to realized that when
the Lorenz number is found to vary significantly with temperature,
so do both electrical and thermal conductivities. This is generic to 
the model in which quasiparticles are responsible for the transport.
Such a model cannot
explain 
experiments\cite{hill} in which the D.C. conductivity is almost
independent of temperature while 
the Lorenz number is 
strongly dependent on it. 

\begin{acknowledgments}
This work was supported in part by the Natural Sciences and Engineering
Research Council of Canada (NSERC) and by the Canadian Institute for Advanced
Research (CIAR).
\end{acknowledgments}

\bibliographystyle{prb}

%
%

\end{document}